\documentclass[12pt,letterpaper]{article}
\usepackage{graphicx}
\usepackage[sort&compress]{natbib}
\usepackage[usenames]{color}
\bibpunct{[}{]}{,}{a}{,}{,}
\newcommand{\noi}{\noindent}
\newcommand{\np}{\newpage\noindent}

\newcommand{\ns}{\normalsize}

\newcommand{\Ls}{\noindent\Large}

\newcommand{\lm}{\lambda}

\newcommand{\eps}{\epsilon}

\newcommand{\rar}{\rightarrow}

\begin{document}
\begin{center}
\vspace*{1in}
\Ls {\bf Maximum Entropy, Time Series and\\ Statistical Inference}\ns
\vspace{1in}

                R. Kariotis   \\
         {\em Department of Physics \\
           University of Wisconsin \\
           Madison, Wisconsin 53706} \\
\end{center}

{\bf ABSTRACT:}
A brief discussion is given of the traditional version of
the Maximum Entropy Method, including a review of some of the criticism
that has been made in regard to its use in statistical inference.
Motivated by these questions,
a modified version of the method is then proposed and applied
to an  example in order to demonstrate its use with a given time series.
\\
\\
\\
\\
\vspace{3in}
\\
e-mail: bobk@physics.wisc.edu
\np
{\bf 1. Historical Background}

The concept of entropy has gone through several distinct stages
since its inception in the 19th century. Originally formulated as
a thermodynamic potential,  Boltzmann reexpressed it as a
measure of disorder. The connection
between thermodynamics and disorder, or randomness, is given by
Boltzmann's ansatz $S=\log(W)$ where $W$ is the count of accessible
configurations
\[
                   W=\frac{N!}{\prod_{i}n_{i}!}
\]
and which leads to the expression
\[
             S=-\sum_{i}p_{i}\log(p_{i})
\]
where $p_{i}$ is the Boltzmann factor for the $i^{th}$ energy state.
In the 20th century, through the work of Shannon [1] and Wiener [2] a second
application was developed where this same expression came to be used as
a measure of the amount of information
contained in a string of characters each of frequency or probability $p_{i}$.

The conceptual association of disorder and information as opposites
was a natural one to make, but
there is considerable disagreement as to how these two ideas,
thermodynamic entropy and information, are related.
Beginning with
Szilard [3], a direct connection between entropy decrease and
information gain was made suggesting that the two differ only by a sign.
Recently Zurek [4] has proposed a more complicated relation suggesting
that gain in information is not entirely reflected in the loss in entropy.
We will come back to the question information and entropy later.
\\
\noi
{\bf 2a. Maximum Entropy and Inference}

Dating from his paper of 1957 [5], ET Jaynes presented a third
use for the expression for entropy, employing it as a tool for
statistical inference.
Consider the problem of constructing the
probability distribution of a system when we have
obtained data/measurements in the form of a set of $M$ numerical values
$\{x_{i}\}$.
With the above definition of entropy and a constraint
on the extremum of the form
\[
             m_{1}= \sum_{i}p_{i}x_{i}
\]
(where $m_{1}$ the first moment) then the full
entropy might be written
\[
             S=-\sum_{i}p_{i}\log(p_{i}) +\lm \sum_{i}p_{i}x_{i}
\]
with $\lm$ the Lagrange multiplier,
and calculating $\delta S=0$ we obtain
\[
  p_{i}=\frac{e^{\lm x_{i}}}{Z} ,    Z = \sum_{i} e^{\lm x_{i}}
\]
The undetermined multiplier is obtained from
\begin{equation}
                 m_{1}=\frac{\partial\log(Z)}{\partial \lm}.
\end{equation}
This set of formulae is directly
analogous to the steps taken in statistical mechanics that relate
 temperature to
the multiplier in equilibrium physics.
The algorithm for constructing the distribution using data is
apparently straight forward, but as it turns out,
there have been a number of
critics over the years who have questioned Jaynes' formalism.
\\
\noi
{\bf 2b. Criticism of the Maximum Entropy Formulation}

In the traditional ME formulation a strict analogy is made between the
microscopic presentation of statistical mechanics, of which
the Bolzmann factor is the principle result, and statistical
inference.
As Jaynes would have it, this is exactly what one
must do for statistical inference; given a time series
whose average is known, one must maximize
\[
             S=\sum_{i}p_{i}\log(p_{i})+\lm \sum_{i}p_{i}x_{i}
\]
where the time series of values $x_{j}$ has been replaced by
\begin{equation}
           \sum_{j}^{M}x_{j} \rar \sum_{i}^{N}n_{i}x_{i}.
\end{equation}
In this expression,
the $M$ observed values of $x_{j}$ are tabulated
and then expressed as a sum of $n_{1}$ occurances of $x_{1}$,
 $n_{2}$ occurances of $x_{2}$, $n_{3}$ occurances of $x_{3}$,....
There are several things wrong with this:\\
\begin{tabbing}
\hspace*{1 in}\=1.\= It is necessary to make a replacement of frequencies\\
\>  \> with probabilities as in eq 2, which contradicts the position that\\
\>  \>  probabilities need not always be associated with frequencies.\\
\>2.  \>  In the variation of $S$ in statistical mechanics, the constraint is\\
\>  \> made up of variables $n_{i}$ over which the variation is taken;\\
\> \> in ME, the constraint is made up of data values,\\
\>  \> which are not free to be varied\\
\>3. \>  There have been several additional criticisms of the ME method\\
\>  \>  beginning with Friedman and Shimony [4,5,6,7]. Good reviews of \\
\>  \>  these questions can be found in Uffink [7]. In particluar,\\
\>  \>  FS find an inconsistancy in ME with respect to its Bayesian \\
\>  \>  properties.  It will  be shown below that the ME formalism will\\
\>  \>  not always reproduce the mean value of the data correctly, and will \\
\>  \>  generally produce a different standard deviation. 
\\
\end{tabbing}

\noi
{\bf 2c. Modified Maximum Entropy}

If we confine the theory to time series data only, a corrected version of ME
 is easily obtained.
Consider the following prescription for statistical
inference.
Given a time series $x_{j}$ of $M$ values we consider the total
number of configurations as
\[
                   W=\frac{N!}{\prod_{i}n_{i}!}
\]
and then weight this product by an extra factor
\[
                   W \rar \frac{N!}{\prod_{i}n_{i}!}Q
\]
where
\[
                 Q= p^{m_{1}}(x_{1})p^{m_{2}}(x_{2})...p^{m_{N}}(x_{N})
\]
which is the probability of a particular string of observations
$\{x_{j}\}$.
That is, $m_{i}$ is the number of times $x_{i}$ has appeared in the time series.
Then the maximization leads to
\[
              \delta S=0=\sum_{i}\delta S_{i}
\]
where each contribution to the sum is of the form
\[
             S_{i}=-p_{i}\log(p_{i})+m_{i}\log(p_{i})
\]
and where the first term (the entropy part) arises from the count of
configurations $W$,
and the last term (the informational part) from the weight $Q$.
The minimization procedure
results in an expression for the probability
(disregarding normalization)
\begin{equation}
               p_{i}=exp(\frac{m_{i}}{p_{i}})
\end{equation}
which may be solved iteratively for $p_{i}$, yeilding
\[
               p(n) =\frac{n}{Z(n)}
\]
where
\[
               Z(n) =\log(n)-\log(\log(n))+
                      {\cal O}\left[\frac{\log(\log(n))}{\log(n)}\right].
\]
Generally,
in the case that successive elements of the time series cannot be
treated as independent, we must use the above equation to solve for
$p(n_{1},n_{2})$, $p(n_{1},n_{2},n_{3})$, etc., depending on the nature
of the correlations in the time series. The evaluation of $W$ in this case
is somewhat more involved, but that of $Q$ remains the same.
\\
\\
{\bf 3. Example Application}\\
\noi
{\bf 3a. Traditional ME}

One of the obvious failings of ME is that it does not reflect the
different contribution made by a short time series versus that of
a long one.
In particular, if the data consisted of a single coin toss, there
does not appear to be a formal solution of the evaluation of the
Lagrange multiplier, i.e. equation 1.
The following is an example of this difficulty (see also Uffink 1996).

Suppose we have a three sided coin $x_{i}=1,2,3$ and toss it
several times with the result $m_{1}$.
This is the first few elements in a time series
and ought to provide a first estimate in the probabilities of the
three sides. According to the above algorithm, the Lagrange
multiplier for this problem is
\[
              m_{1}=\frac{x_{1}q+x_{2}q^{2}+x_{3}q^{3}}{Z}
\]
where $q=e^{-\lm}$. The prescription is to solve for $\lm$, that is
\begin{equation}
        (m_{1}-x_{1})q+(m_{1}-x_{2})q^{2}+(m_{1}-x_{3})q^{3}=0
\end{equation}
A plot of the largest real positive root is given in Figure 1.
If the time series contains only a few elements, it is possible
that the average turns out to be $m_{1}=1$ or $m_{1}=3$
yet according to the plot this requires $q=0$ or $q=\infty$,
 i.e. no disorder regardless of the data.\\

\begin{figure}[!h]
\begin{center}
\includegraphics[width=2.5in,height=2in]{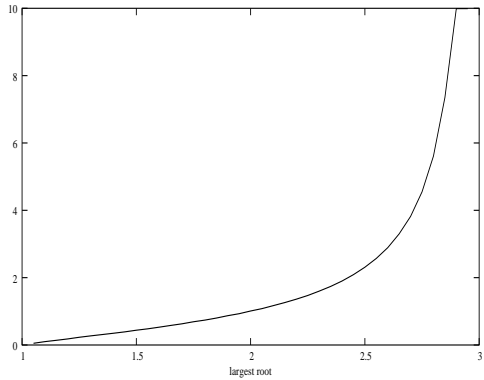}
\caption{the root of equation 4 with respect to mean value}
\end{center}
\end{figure}

This difficulty extends to the moments in general.
For example, for a two level system with ''energies'' $\eps_{1},\eps_{2}$,
in Jaynes' formulation, moments are determined from data
\[
                       m_{l}=\sum_{j}x^{l}_{j}
\]
and these in turn are related to the Lagrange multipliers by
\[
                         m_{l}=\frac{1}{Z}\left[\eps_{1}^{l}e^{\phi_{1}}+\eps_{2}^{l}e^{\phi_{2}}\right]
\]
with
\[
               Z=e^{\phi_{1}}+e^{\phi_{2}}
\]
where the multipliers are given by
\[
     \phi_{i}=\lm_{1}\eps_{i}+\lm_{2}\eps_{i}^{2}+\lm_{3}\eps_{i}^{3}+...
\]
For the two-level system, this works out to be
\[
                          -\frac{(m_{l}-\eps_{1}^{l})}{(m_{l}-\eps_{2}^{l})}
                                       =e^{\phi_{2}-\phi_{1}}
\]
for all $l$, which is overdetermined in general and so has no solution.
\\
\\
\\
\noi
{\bf 3b. Modified ME}

In contrast the modified method yields probability values for any length
of time series, and in the limit of a large string, gives the same
value as would be obtained from a direct frequency tabulation.
As a demonstration of how this operates, consider the toss
of a two-sided coin, using a possibly biased coin. After
$N=n_{h}+n_{t}$ tosses the probability for each side is given by
$P_{h}=\frac{p(n_{h})}{p(n_{h})+p(n_{t})}$ 
and 
$P_{t}=\frac{p(n_{t})}{p(n_{h})+p(n_{t})}$,
where $p(n_{h})$  is the solution to equation 3 above.

 Normalization of the probabilities is given by
\[
                  Z=p_{h}(n_{h})+p_{t}(n_{t})
\]
The
case similar to the one described above by Uffink is where a single
toss results in a head which is obtained from
\[
                  Z=p_{h}(1)+p_{t}(0)
\]
In Figure 2 we plot the probability of heads and tails respectively, in
a simulation where the coin is biased such that in the infinite limit
we should have $p_{h}=.7$ and $p_{t}=.3$.
\begin{figure}[!ht]
\begin{center}
\includegraphics[width=2.5in,height=2in]{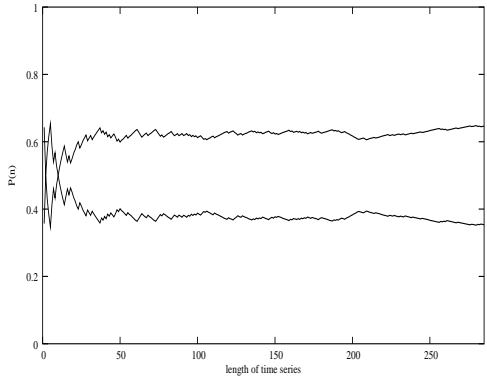}
\caption{the probabilities for the biased coin as a function
             of the number of tosses}
\end{center}
\end{figure}
As can be seen the initial
values are rough, while in the limit of large data the values go
over into the frequency values.

Returning now to the problem discussed above, consider a 2-sided coin
that can only land heads up, such that  successive throws yield
an increasingly long string of heads. 
Regardless of the string length,
the traditional ME can only
give the solution $(p_{h},p_{t})=(1,0)$, as shown in Section 3a,
while the modified version gives $p_{t}=1-p_{h}$ and
\[
                       p_{h}=\frac{p(n)}{1+p(n)}
\]
after the $n^{th}$ throw. This is plotted in Figure 3 and provides
a more realistic description of how an experimenter might gradually,
but inevitably, come to the conclusion that the toss is not random.
\begin{figure}[!ht]
\begin{center}
\includegraphics[width=2.5in,height=2in]{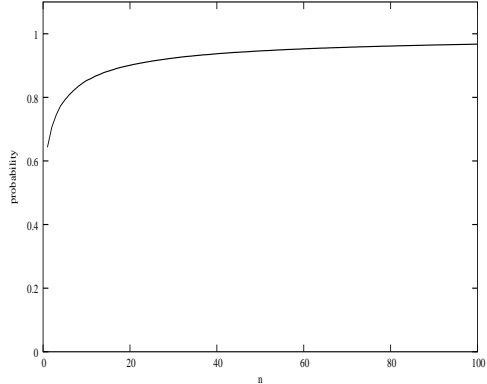}
\caption{probability of heads for the completely biased coin}
\end{center}
\end{figure}
\\
\\
\\
\\
\noi
{\bf 4. Discussion}

In summary we have proposed that the traditional method of applying
the Maximum Entropy method, which involves maximization of
\[
                  S= -\sum_{i}p_{i}\log(p_{i})+\lm\sum_{i}p_{i}x_{i}
\]
must be modified to
\[
                  S= -\sum_{i}p_{i}\log(p_{i})+ \sum_{i}m_{i}\log(p_{i})
\]
when dealing with time series.
This allows the elements of a time series, no matter how few in number, to
be included and hence influence the resulting distribution.
This method has several advantages, most notably that the elements of
the time series can be put into the computation without difficulty
and without complication whether the series is short or long.
Also, as the series becomes very long, the probability values
become equal to those found from direct frequency tabulation.
One undeniable disadvanteage is that the resulting distribution is
not usually a smooth function of the data.  That is,
for a short series,
a small number of new data points may make a considerable change in
the numerical values of the $p_{i}$; this is in contrast with the
traditional ME approach,
but perhaps is a more realistic property of statistical inference.

The modified expression is made up of an entropy component and an
informational one. For short time series with little or no data,
the entropy contribution dominates,
starting off with the equal probability distribution when no data at all
is at hand.
Then, as data is accumulated (the longer time series),
the informational part dominates,
and it is to be expected that the manner of cross-over from one
to the other should depend on the nature of the data.
The interpretation of this is easily seen by considering the example
of the completely biased coin where $p_{h}=1$.
The information obtained from a single toss is given by
$p_{h}\log(p_{h})+ p_{t}\log(p_{t})=0$; 
this
makes sense as the outcome is already known; however, we might ask how much
information had been acquired in determining the distribution originally?
The results obtained here suggest that the additional
term in the expression for entropy provides a means of fixing a value to
the information contained in the statement of the distribution. 
These ideas are  somewhat in line with earlier
studies (Zurek [4], Lin [10]), 
though the context of this added term is different from these papers.
In order to find how much information is already
contained in the knowledge of the distribution, we must consider the
manner in which the distribution was obtained. Initially we might
start with $p_{h}=p_{t}=\frac{1}{2}$, and begin tossing, keeping
record of the results. Each toss modifies $p_{h}$ and $p_{t}$
according to the function $p(n)$ given in equation 3.
As the number of tosses grows indefinitely, the amount of information
contained in the data is given by the limiting value shown above,
$-n\log(p)$, which for the completely biased coin leads to $\log(n)$
for large $n$.
\np
{\bf Acknowledgment:} I want to thank A. Shimony for making comments on an
earlier version of this paper, and I also want to thank L. Bruch for his
continued assistance and many offerings of good advice.
\\
\\
\\
{\bf References:}\\
1. C Shannon, A Mathematical Theory of Communication; U Ill Press (1949)\\
2. N Wiener, Cybernetics; Wiley and Sons  (1948)\\
3. L Szilard, reprinted in H Leff and A Rex, Maxwell's Demon, Princeton (1990)
\\
4. W J Zureck Phys Rev A \underline{40} 4731 (1989)\\
5. E T Jaynes, Maximum Entropy Formalism, p.15, RD Levine,M Tribus (eds) MIT//
\hspace*{.5in}(1979); E T Jaynes, IEEE Proceedings  \underline{70} 939 (1982);\\
\hspace*{.5in} E T Jaynes,  Phys Rev  \underline{106} 620 (1957)\\
6. K Friedman and A Shimony,  J Stat Phys  \underline{3} 381 (1971)\\
7. T Seidenfeld, Phil Sci  \underline{53}  467 (1986)\\
8. A Shimony,  Synthese  \underline{63} 35 (1985);
 reprinted in A. Shimony,\\
\hspace*{.5in} Search for a Naturalistic World View,  Cambridge (1993)\\
9. J Uffink, Stud Hist Phi Mod Phys  \underline{26} 223 (1995);
 Stud Hist Phi Mod Phys \underline{27} 47 (1996)\\
10. S K Lin Int J Mol Sci \underline{2} 10 (2001)\\
\end{document}